\documentclass[lettersize,journal]{IEEEtran}

\usepackage{soul,color}
\usepackage{amsmath,amsfonts}
\usepackage{algorithmic}
\usepackage{algorithm}
\usepackage{array}
\usepackage{textcomp}
\usepackage{stfloats}
\usepackage{url}
\usepackage{verbatim}
\usepackage{graphicx}
\usepackage{cite}
\usepackage{multirow}
\usepackage{float}
\usepackage{tipa}
\usepackage{subcaption}
\usepackage{arydshln}
\usepackage{hyperref}
\usepackage{caption}
\begin{document}

\title{Leveraging Laryngograph Data for Robust Voicing Detection in Speech}
%
\author{Yixuan Zhang, Heming Wang,~\IEEEmembership{Student Member,~IEEE,} DeLiang Wang,~\IEEEmembership{Fellow,~IEEE,}
\thanks{Y. Zhang is with the Department of Computer Science and Engineering, The
Ohio State University, Columbus, OH 43210 USA (e-mail: zhang.7388@osu.edu).}
\thanks{H. Wang is with the Department of Computer Science and Engineering, The
Ohio State University, Columbus, OH 43210 USA (e-mail: wang.11401@osu.edu).}
\thanks{D. L. Wang is with the Department of Computer Science and Engineering
and the Center for Cognitive and Brain Sciences, The Ohio State University,
Columbus, OH 43210 USA. (e-mail: dwang@cse.ohio-state.edu).}}

%
%

%
\maketitle
\begin{abstract}

Accurately detecting voiced intervals in speech signals is a critical step in pitch tracking and has numerous applications. While conventional signal processing methods and deep learning algorithms have been proposed for this task, their need to fine-tune threshold parameters for different datasets and limited generalization restrict their utility in real-world applications. To address these challenges, this study proposes a supervised voicing detection model that leverages recorded laryngograph data. The model is based on a densely-connected convolutional recurrent neural network (DC-CRN), and trained on data with reference voicing decisions extracted from laryngograph data sets. Pretraining is also investigated to improve the generalization ability of the model. The proposed model produces robust voicing detection results, outperforming other strong baseline methods, and generalizes well to unseen datasets. The source code of the proposed model with pretraining is provided along with the list of used laryngograph datasets to facilitate further research in this area.
\end{abstract}
%

\section{Introduction}
\label{sec:intro}
A speech signal consists of voiced, unvoiced, and silent intervals or segments. Detecting whether speech is voiced is known as voicing detection. 
This task is a crucial step in pitch estimation and benefits various speech processing tasks, such as speaker recognition \cite{bai2021speaker}, computational auditory scene analysis \cite{wang2006computational}, and speech recognition \cite{zolnay2002robust, atal1976pattern}. 
In deep learning, precise voicing detection contributes to enriching training data with essential context and segmentation. This can be especially valuable in situations where annotated data is limited. With voicing information, the potential of deep learning in advancing speech processing tasks can be further leveraged.
It is important to note that voicing detection is different from voice activity detection (VAD), which aims to determine the presence or absence of speech activity in an audio signal. In contrast, voicing detection concerns detecting the voiced portions of speech signals.

Voiced speech is produced by the vibration of the glottis, creating periodic or semi-periodic pulses of air that resonate through the vocal tract, while unvoiced speech is aperiodic and produced when air flows through a narrow constriction in a way to produce turbulence noise with no glottis vibration \cite{stevens1998acoustic}. Therefore, periodicity is the determining factor for voiced and unvoiced segments. In English, voiced sounds include all vowels and voiced consonants such as /g/, /v/, and /z/, while unvoiced sounds include unvoiced consonants such as fricatives (e.g. /f/) and stops (e.g. /p/). Unvoiced speech accounts for approximately 20-25\% of all speech sounds in terms of both phoneme occurrence and segment duration \cite{hu2008segregation}, which highlights the significant role that unvoiced sounds play in speech utterances. Further description of voiced and unvoiced speech segments in English will be provided in Sec. \ref{sec:voicing}.

Various approaches have been proposed to address voicing detection, by analyzing the waveform or energy of the signal, or by examining spectral characteristics such as the presence of harmonics and formants. Conventional methods include analyzing the short-term autocorrelation sequence \cite{de2002yin}, zero-crossing rate \cite{amado2008pitch}, and energy of the speech signal \cite{bachu2010voiced}, as well as a combination of these features with classification techniques like thresholding \cite{talkin1995robust} and rule-based approaches \cite{wang2022robust}.  
Deep learning based approaches for voicing detection often treat the task as part of pitch tracking and train a multi-task model \cite{han2014neural, tran2020robust, zhang2022densely, morrison2023cross}. Pitch contours obtained from laryngograph data are commonly used as the ground-truth for evaluating the performance of these algorithms.
However, existing methods for voicing detection have limitations in robustness and generalizability. Signal processing methods are sensitive to various types of noise, including device noise, and typically require ad-hoc tuning of a voicing decision threshold for each dataset, i.e. lacking consistency \cite{amado2008pitch, bachu2010voiced, talkin1995robust}. 
While deep neural network (DNN) methods \cite{kim2018crepe, morrison2023cross} can perform voicing detection on clean speech by applying a threshold to the estimated periodicity, the sensitivity of this threshold still presents a challenge. The optimal threshold can vary depending on the dataset used. Furthermore, the limited training data can lead to generalization issues, resulting in unreliable performance.

To address the above issues with existing methods, we propose to train a voicing detection model with multiple datasets that provide laryngograph recordings. Specifically, we directly train a DC-CRN (densely-connected convolutional recurrent network) model using ground-truth labels extracted from the laryngograph data. Voicing labels obtained from laryngograph recordings are generally considered the gold standard for evaluating the accuracy of voicing detection models. By gathering existing datasets that contain laryngograph recordings, we obtain an adequate amount of data to train the proposed model.

Laryngograph, also known as electroglottograph, is a medical device used for measuring the electrical activity of the larynx during speech production. It is a non-invasive device that is placed on the skin of the neck and detects changes in the electrical impedance of the vocal folds as they vibrate by emitting a high-frequency electrical signal into the neck. These changes in impedance are used to create a waveform that represents the movement of the vocal folds during pronunciation. 
Compared to microphone recordings, laryngograph recordings have several advantages for producing accurate voicing decisions. First, it provides a direct measure of the vibration of the vocal folds, which is the source of the voiced speech signal. This is more accurate than methods that rely on indirect measures of voicing, such as the spectral or temporal characteristics of the speech signal.
Second, the laryngograph is relatively unaffected by variations in the amplitude or frequency of the speech signal caused by acoustic noise or interference, which can be a problem for voicing detection methods based on microphone recordings. This makes the laryngograph a reliable tool for detecting voicing, especially in adverse acoustic environments.
On the other hand, conducting laryngograph recordings is a cumbersome job. As a result, publicly accessible laryngograph data is limited, particularly from the perspective of large-scale DNN training.   

This paper presents a robust voicing detection model for clean speech that achieves state-of-the-art performance by leveraging multiple laryngograph datasets for training. We find that the model trained on accessible laryngograph datasets already yields good generalization. To further mitigate potential generalization issues, we conduct pretraining on the large-scale Librispeech dataset \cite{panayotov2015librispeech}, which leads to improved and more robust voicing detection performance. The contributions of our work can be summarized as follows:
\begin{itemize}
    \item We investigate the distinct characteristics of voiced and unvoiced speech sounds, and assess the feasibility of training on laryngograph datasets.
    \item We develop a robust supervised voicing detector capable of accurately estimating voicing in clean speech.
    \item We propose to use pretraining to further enhance the accuracy of voicing detection.
    \item We conduct a comprehensive evaluation of the proposed method and comparing against other strong baselines.
    \item We release a pip-installable Python library containing the trained model, which can be used to generate reliable ground-truth labels in cases where laryngograph data is not available, along with compiled laryngograph datasets.
\end{itemize}


This paper is structured as follows. In Sec. \ref{sec:voicing}, we provide a description of voiced and unvoiced English speech sounds and their characteristics. Sec. \ref{sec:related} and Sec. \ref{sec:datasets} describe related works and publicly accessible laryngograph datasets. Sec. \ref{sec:method} presents our voicing detection model. Sec. \ref{sec:setup} and Sec. \ref{sec:eval} describe our experimental setup and evaluations of the proposed approach, including comparisons with existing methods. Finally, Sec. \ref{sec:conclusions} provides concluding remarks. The source code and pre-trained model used in this study are provided at \href{https://github.com/YIXUANZ/rvd}{https://github.com/YIXUANZ/rvd}.

\section{Voiced and Unvoiced Speech Sounds}
\label{sec:voicing}

\begin{table}[!t]
\centering
\caption{Voiced and unvoiced phonemes in English.}
\label{tbl:phrases}
\begin{tabular}{c|c|c}
\hline
\textbf{Phoneme Type}                            &       \textbf{Phoneme}                                             & \textbf{Voiced or Unvoiced?} \\ \hline
Vowels                      & All                                                      & Voiced             \\ \hline
Approximants                & All                                                      & Voiced             \\ \hline
Nasals                      & All                                                      & Voiced             \\ \hline
\multirow{2}{*}{Stops}      & /d/, /b/, /g/                                            & Voiced             \\ \cdashline{2-3}
                            & /t/, /p/, /k/                                            & Unvoiced           \\\hline
\multirow{2}{*}{Fricatives} & /z/, /v/, /\textipa{Z}/, /\textipa{D}/                                 & Voiced             \\  \cdashline{2-3}
                            & /s/, /f/, /\textipa{S}/, /\textipa{T}/ & Unvoiced           \\ \cdashline{2-3}
                            & /h/ & Both\\\hline
\multirow{2}{*}{Affricates} &  /d\textipa{Z}/                                                        & Voiced             \\ \cdashline{2-3}
                            & /t\textipa{S}/                                                         & Unvoiced           \\ \hline
\end{tabular}
\end{table}


How to distinguish between voiced and unvoiced speech sounds? As discussed in Sec. \ref{sec:intro}, the primary characteristic is periodicity, which is evident as harmonic patterns in the frequency domain. Therefore, detecting frames with harmonic patterns in the spectrum of the speech signal becomes an intuitive approach. Nonetheless, this task can be difficult in certain scenarios.
Unvoiced frames exhibit no harmonic patterns in their spectrum, and can be challenging to distinguish from background noise. Although the presence of harmonic structure is a reliable indicator of a voiced frame, it can be still difficult to recognize such harmonic patterns in frames located between voiced and unvoiced intervals due to co-articulation effects. In such cases, harmonic patterns may be ambiguous, even though harmonic components still exist in the signal. In order to characterize these ambiguous frames, one can utilize contextual cues to make a determination. Linguistic features of the language being spoken provide helpful clues in distinguishing voiced and unvoiced speech segments. In English, phonemes are classified as either voiced or unvoiced \cite{ladefoged2001vowels}. Table \ref{tbl:phrases} provides a catalog of the voiced and unvoiced phonemes in English, where all vowels, approximants, and nasals are voiced \cite{ladefoged2001vowels}. Certain consonants, including stops, fricatives, and affricates, have pairs of voiced or unvoiced sounds. It should be noted that the phoneme /h/ can be pronounced either in a voiced or unvoiced way.

The use of a laryngograph provides an effective way to distinguish between voiced and unvoiced frames empirically. 
As explained in Sec. \ref{sec:intro}, laryngograph recordings provide a direct measurement of the vibrations from the source of the voiced signal, which is relatively unaffected by amplitude or frequency variations caused by environmental noise or interference.

\begin{figure}[t]
  \centering
  \includegraphics[width=\linewidth]{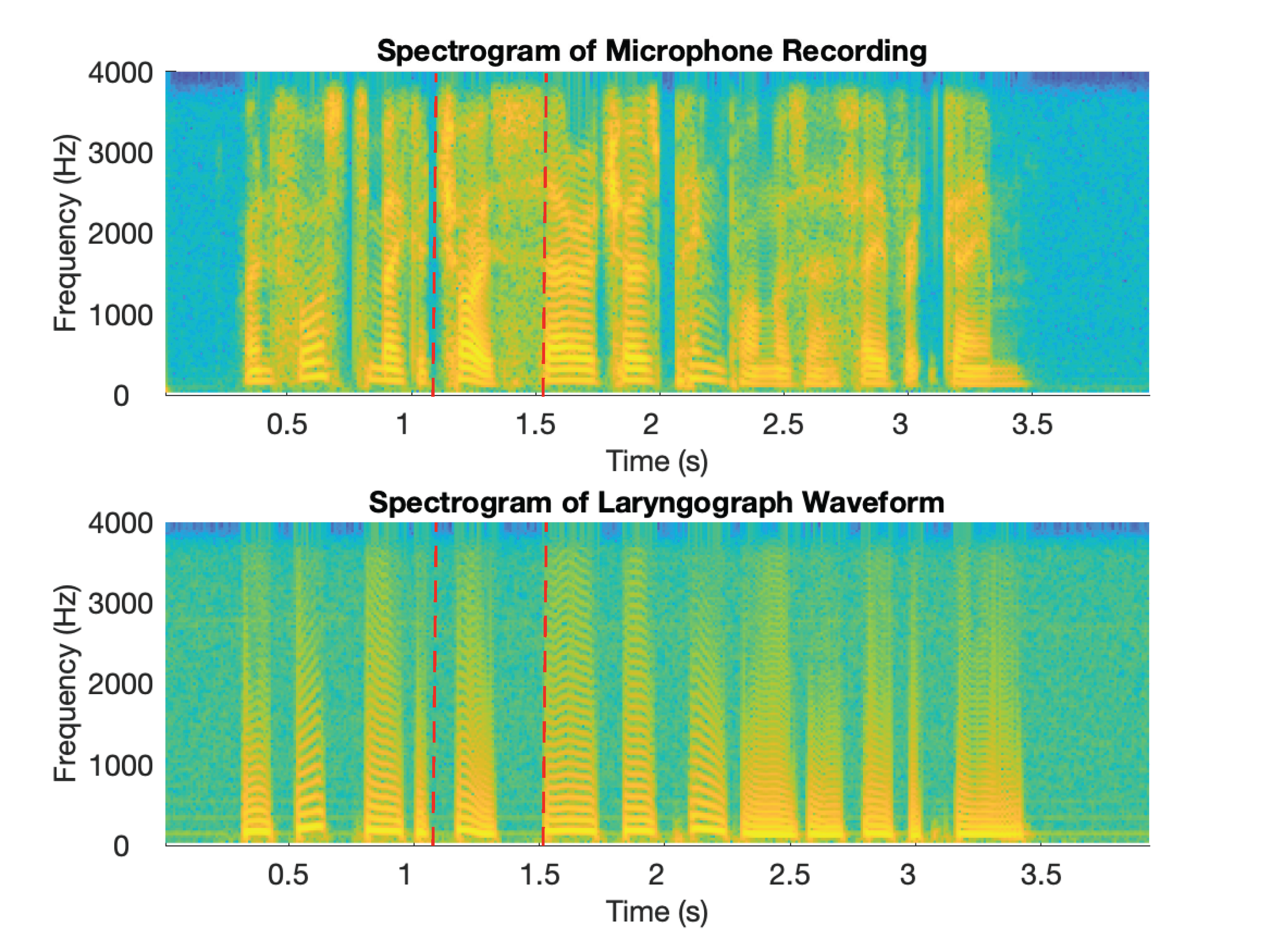}
  \caption{Magnitude spectrograms of microphone and laryngograph recordings of an utterance from a female speaker in the FDA dataset \cite{bagshaw1993enhanced}. The demarcated interval corresponds to the word ``choice".}
  \label{fig:comparison}
\end{figure}

Figure \ref{fig:comparison} shows a comparison between the magnitude spectrogram of a microphone recording and a laryngograph waveform. The audio recording is from the FDA dataset \cite{bagshaw1993enhanced} and corresponds to the utterance ``When forced to make a choice, Sarah chose ping-pong as her favorite game." We can observe that the laryngograph spectrogram provides a clear distinction between voiced and unvoiced intervals of the speech signal. For example, the word ``choice" can be observed in the spectrograms between 1.122 and 1.537 s. This word is composed of both voiced and unvoiced sounds, but the unvoiced sounds are not captured in the laryngograph. The microphone recording, however, exhibits both kinds of sound, potentially complicating voicing detection.

\section{Related Works}
\label{sec:related}

Numerous studies have been conducted for voicing detection given its importance for applications such as speech synthesis. Earlier studies primarily focus on developing signal processing algorithms \cite{drugman2019joint, talkin1995robust, narendra2015robust, koutrouvelis2015fast, upadhyay2015instantaneous, kumar2016voice, mcaulay1990pitch}. Among these algorithms, the robust algorithm for pitch tracking (RAPT) \cite{talkin1995robust} and the summation of residual harmonics (SRH) \cite{drugman2019joint} algorithm are considered as the standard methods in clean and noisy speech, respectively \cite{koutrouvelis2015fast}. RAPT is a time-domain method that employs the normalized cross-correlation function (NCCF) \cite{atal1972automatic} and dynamic programming for pitch tracking. In the post-processing stage, a voicing decision is made by applying dynamic programming to select the set of NCCF peaks in frames containing voiced speech signals, or to make no selection otherwise. SRH \cite{drugman2019joint} leverages harmonic information in the residual signal to estimate pitch and make voicing decisions. It calculates SRH using the amplitude spectrum of the residual signal. During unvoiced intervals of speech, SRH values tend to be lower. Therefore, the algorithm applies a simple local threshold to SRH values to make voicing decisions, and a speech frame is classified as voiced if its SRH value is above the threshold, and unvoiced otherwise.

In recent years, there has been a growing interest in exploring deep learning approaches for voicing detection, but primarily focusing on noisy or multi-talker scenarios. These approaches aim to address voicing detection and pitch estimation simultaneously. For example, studies in \cite{han2014neural, liu2018permutation} treat the two tasks as a multi-class classification problem, while others \cite{tran2020robust} employ a multi-task learning approach to jointly perform the two tasks. In these methods, ground-truth labels are obtained by applying a pitch tracker to the microphone recordings of clean speech, which limits the accuracy of the trained model due to errors introduced by the pitch tracker on such clean speech. When it comes to ground truth, as discussed earlier, laryngograph data is considered to be the most reliable reference \cite{plante1995pitch}. Several approaches have been proposed to address voicing detection in clean speech. Among them is the representative CREPE \cite{kim2018crepe}, which is trained on synthetic data to provide a periodicity estimate for each frame. The degree of periodicity indicates the likelihood of the presence of voiced speech within the frame, with higher values indicating a greater likelihood of voicing. Similar to SRH, a simple threshold is utilized to determine whether a frame is voiced. Such an approach sometimes produces unreliable voicing decisions. In \cite{morrison2023cross}, an entropy-based method is introduced for generating periodicity, and along with several training strategies, it significantly improves the accuracy of voicing decisions. Another approach involves utilizing a laryngograph to create annotations, which can then be employed to train a model on microphone recordings. For example, Drugman et al. \cite{drugman2018traditional} incorporate both internal data and the CMU Arctic dataset \cite{kominek2004cmu} in their training data. Labels are obtained from laryngograph data and a leave-one-speaker-out cross-validation scheme is employed during training to assess the effectiveness of their approach. While the idea is sensible, there is certainly room for improvement. First, their training dataset is relatively small, which potentially limits the generalizability of their trained model. Second, they employ a plain multi-layer perception (MLP), which may not be able to model complex patterns in the data as well as more advanced deep neural networks. 

To our knowledge, there is currently no open-source DNN-based voicing detection algorithm trained on accessible laryngograph data, which hinders the effort of building on and improving earlier work. Our study intends to rectify this situation.



\section{Laryngograph Datasets, Preprocessing, and Label Generation}
\label{sec:datasets}


\subsection{Laryngograph Datasets}
\label{ssec:laryndata}
\begin{table*}[!t]
\centering
\caption{Description of accessible laryngograph datasets}
\label{tbl:datasetinfo}
\begin{tabular}{c|cccc}
\hline
\textbf{Dataset}     & \textbf{Speaker Information}          & \textbf{\# of Utterances} & \textbf{Label Provided?} & \textbf{Label Extraction Method}   \\ \hline
FDA \cite{bagshaw1993enhanced}        & 1 male and 1 female   & 100              & Yes             & Pulse Location Algorithm  \\
PTDB-TUG \cite{pirker2011pitch}    & 10 male and 10 female & 4720             & Yes             & RAPT Algorithm            \\
KEELE \cite{plante1995pitch}       & 5 male and 5 female   & 98 (approximated)               & Yes             & Autocorrelation Algorithm \\
Mocha-TIMIT \footnote{https://data.cstr.ed.ac.uk/mocha/} & 4 male and 5 female   & 4028             & No              & -                         \\
CMU Arctic \cite{kominek2004cmu} & 2 male and 1 female   & 3377             & No              & -                         \\ \hline
Total       & 22 male and 22 female & 12323            & -               & -                         \\ \hline
\end{tabular}
\end{table*}

Voicing labels generated from laryngograph recordings are widely used as ground-truths for evaluating voicing detection methods. 
Table \ref{tbl:datasetinfo} lists publicly accessible datasets employed in this study, and provides relevant details for each dataset. We do not incorporate publicly accessible datasets that provide fewer than 100 utterances. Among the five datasets, three provide reference pitch and voicing labels extracted by different algorithms. FDA \cite{bagshaw1993enhanced} is a relatively small dataset that provides microphone and laryngograph recordings from a male and a female speaker, and each speaker has 50 utterances. The provided reference labels in the FDA dataset are extracted using a `pulse' location algorithm where the duration between consecutive pulses are derived and converted to Hertz. If the value is within a certain range, the duration is considered voiced. Otherwise, it is considered unvoiced. PTDB-TUG \cite{pirker2011pitch} has 10 male speakers and 10 female speakers and around 4720 utterances in total. The provided reference labels are extracted by first applying a high-pass filter on laryngograph waveforms to remove low frequency components caused by larynx movements and then applying the RAPT \cite{talkin1995robust} algorithm on the filtered laryngograph waveforms. The KEELE \cite{plante1995pitch} dataset has recordings from 5 adult male speakers, 5 adult female speakers, and 5 children. We can only find the recordings from adult speakers. For male speakers, the length of each recording is from 27 seconds to 40 seconds. For female speakers, the length is from 28 to 30 seconds. To better process the data, we further split the recordings to utterances around 3 second-long. In total, we obtained 98 utterances. Mocha-TIMIT \footnote{https://data.cstr.ed.ac.uk/mocha/} and CMU Arctic \cite{kominek2004cmu} are relatively large datasets but do not provide reference labels. The Mocha-TIMIT dataset has 4028 utterances, which are uttered by 4 male and 5 female speakers. In the CMU Arctic dataset, we find that the recordings from 2 male and 1 female speakers come with laryngograph waveforms. In total, the collected datasets contain 12323 utterances from 22 male and 22 female speakers.

\subsection{Data Preprocessing} 
\label{ssec:preprocessing}

\begin{figure}[t]
  \centering
  \subfloat[]{\label{fig:ptdb_err}{\includegraphics[width=\linewidth]{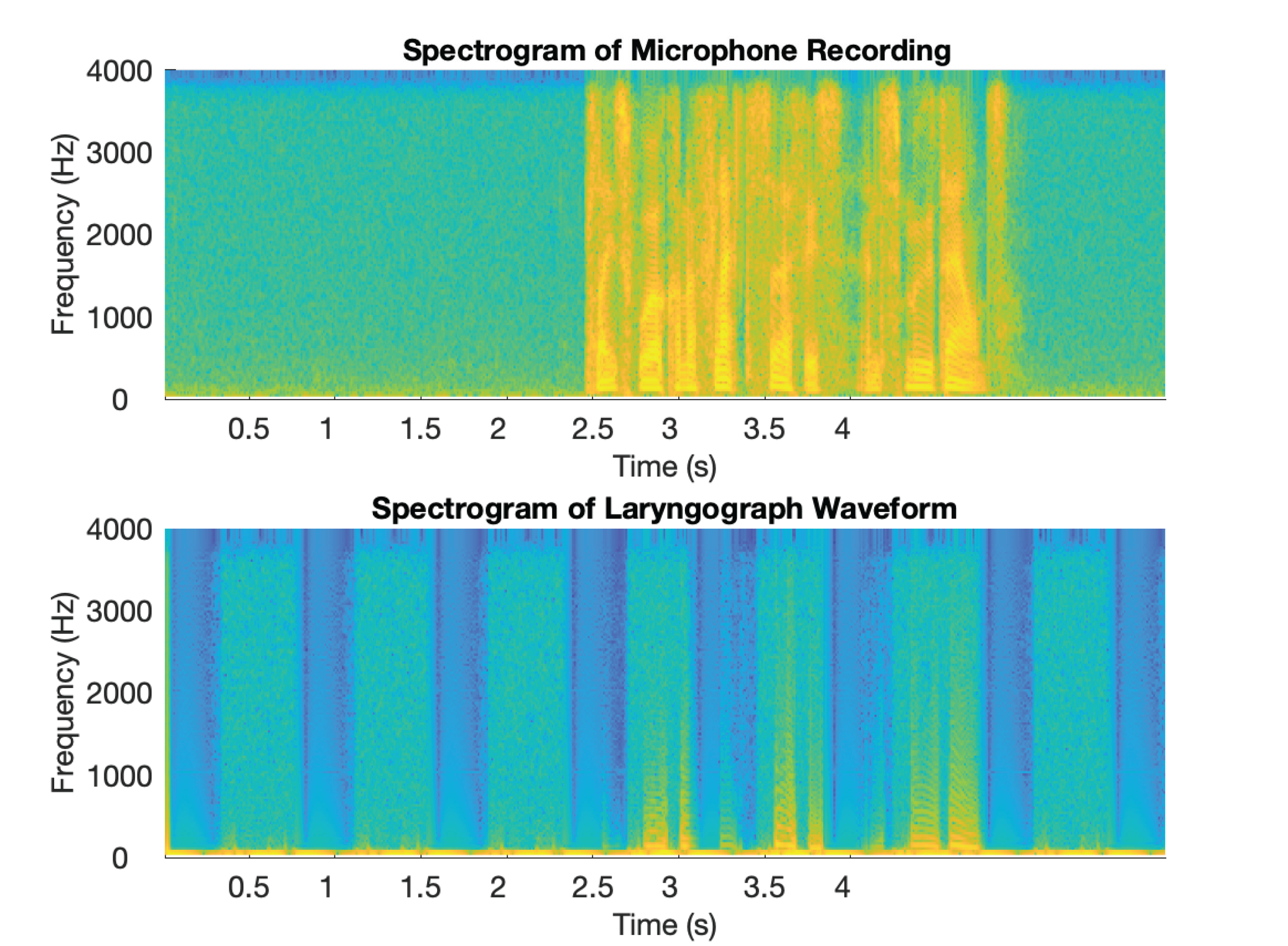}}}\hfill \\
    \subfloat[]{\label{fig:mocha_err}{\includegraphics[width=\linewidth]{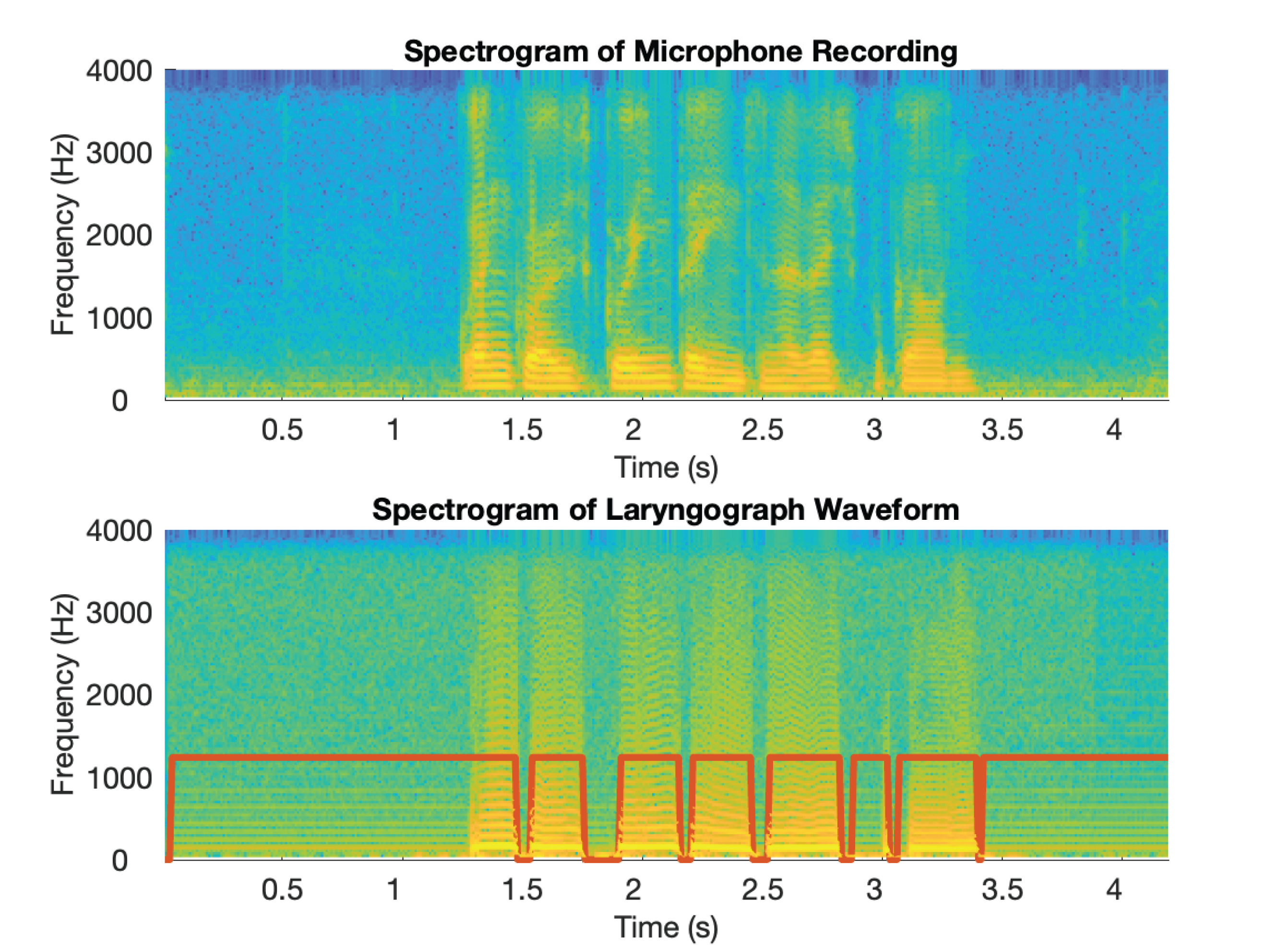}}}\hfill \\
  \caption{Examples of low-quality laryngograph data in (a) PTDB-TUG dataset and (b) Mocha-TIMIT dataset, with corresponding spectrograms of microphone and laryngograph recordings. The red line in (b) represents the reference voicing labels including the erroneous labels extracted from the flawed laryngograph waveform.}
  \label{fig:laryn_err}
\end{figure}

While the laryngograph data in these datasets is generally suitable for training purposes, our review reveals that certain datasets, such as PTDB-TUG, contain flawed laryngograph recordings. Additionally, we observe that some laryngograph data in the Mocha-TIMIT dataset contains noise with a harmonic pattern during periods of silence. Examples of these issues are illustrated in Figure \ref{fig:laryn_err}. Directly using such data for training without resolving these issues will negatively impact the accuracy of model  training and evaluation. Therefore, such utterances should be either excluded or carefully processed. 



Specifically, we find that two of the five datasets in Table \ref{tbl:datasetinfo} - PTDB-TUG and Mocha-TIMIT - contain problematic recordings. 
Upon examining the PTDB-TUG dataset, we find a number of laryngograph waveforms to be of low quality, such as the one depicted in Figure \ref{fig:ptdb_err}. These recordings do not appear to capture vocal fold movements, making them unsuitable for extracting ground-truth labels. We decide to remove these recordings from training or evaluation sets.
For the Mocha-TIMIT dataset, we have identified some files that contain noisy harmonic patterns in silence intervals, as illustrated in Figure \ref{fig:mocha_err}. These silence intervals would be recognized as voiced frames by a pitch extraction algorithm due to the presence of harmonic structure. 
To ensure evaluation accuracy, we manually correct the labels in these intervals of the affected files. Files that pose challenges for correction are omitted.
A total of 1230 waveforms are excluded from the PTDB-TUG dataset, which originally comprises 4720 audio samples. Additionally, nearly 500 waveforms are corrected and around 270 waveforms are excluded for the Mocha-TIMIT dataset.

\subsection{Label Generation}
\label{ssec: label generation}
Given laryngograph data, different algorithms can be used to extract ground-truth voicing labels, and there is no standard way to perform label extraction. Although different algorithms produce similar results, the results differ to some extent. It is common that a paper announcing a laryngograph dataset provides reference labels and encourages users to generate their own reference labels \cite{plante1995pitch, pirker2011pitch}.

In alignment with the method outlined for reference voicing label generation in PTDB-TUG \cite{pirker2011pitch}, we employ the following steps to derive reference voicing labels from laryngograph datasets: 
\begin{itemize}
    \item Pre-process each dataset and manually remove all utterances with quality issues.  
    \item High-pass filter each utterance to remove the lower frequency components caused by larynx movements. Specifically, apply a linear phase Kaiser filter with parameters $\beta = 5$ and $n = 2400$ to laryngograph signals. For female speaker signals the cut-off frequency is set to $f_c=25$ Hz, and for the male speakers $f_c=15$ Hz.
    \item Apply the RAPT algorithm to filtered laryngograph signals to produce voicing decisions.
    \item If an audio frame is considered voiced, the reference label $y_v$ is set to 1. Otherwise, it is set to 0.
\end{itemize}

\begin{table}[]
\centering
\caption{Mismatch rates (\%) between provided labels and self-generated labels}
\label{tbl:labelgen}
\begin{tabular}{c|c}
\hline
\textbf{Dataset}  & \textbf{Mismatch Rate (\%)} \\ \hline
FDA      & 1.89\%             \\
KEELE    & 2.19\%             \\
PTDB-TUG & 1.90\%             \\ \hline
\end{tabular}
\end{table}

\begin{figure}[t]
  \centering
  \includegraphics[width=\linewidth]{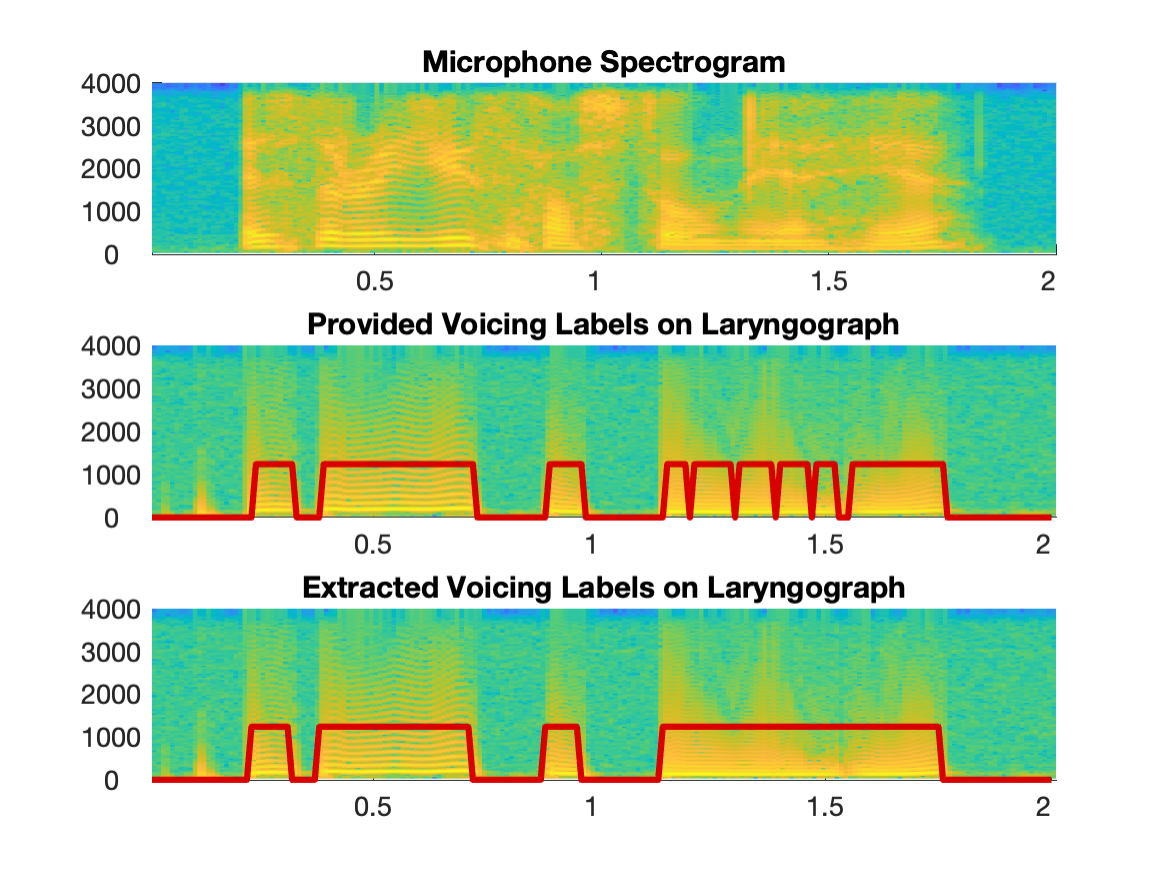}
  \caption{Comparison of provided and extracted voicing labels on the utterance with the highest mismatch rate in the FDA dataset (corresponding to utterance rl006 in the FDA dataset). Red line represents voicing decisions where 0 indicates unvoiced and other positive value indicates voiced.}
  \label{fig:provided_ref}
\end{figure}

How much do different label extraction algorithms differ? We use the above method to extract reference labels from PTDB-TUG, KEELE and FDA datasets and compared to the provided reference labels. Alignment is performed to maximize the match between the provided and extracted labels. The mismatch rates, which represent the percents of mismatched frames to all frames are given in Table \ref{tbl:labelgen}. We observe that the mismatch rate is around 2\% for all datasets. Figure \ref{fig:provided_ref} shows an example utterance with the top mismatch rate in the FDA dataset. The figure shows that the provided reference labels tend to under-label voiced intervals, and our method provides more balanced voicing decisions.

\section{Model Description}
\label{sec:method}

\begin{figure}[t]
  \centering
  \includegraphics[width=0.5\linewidth]{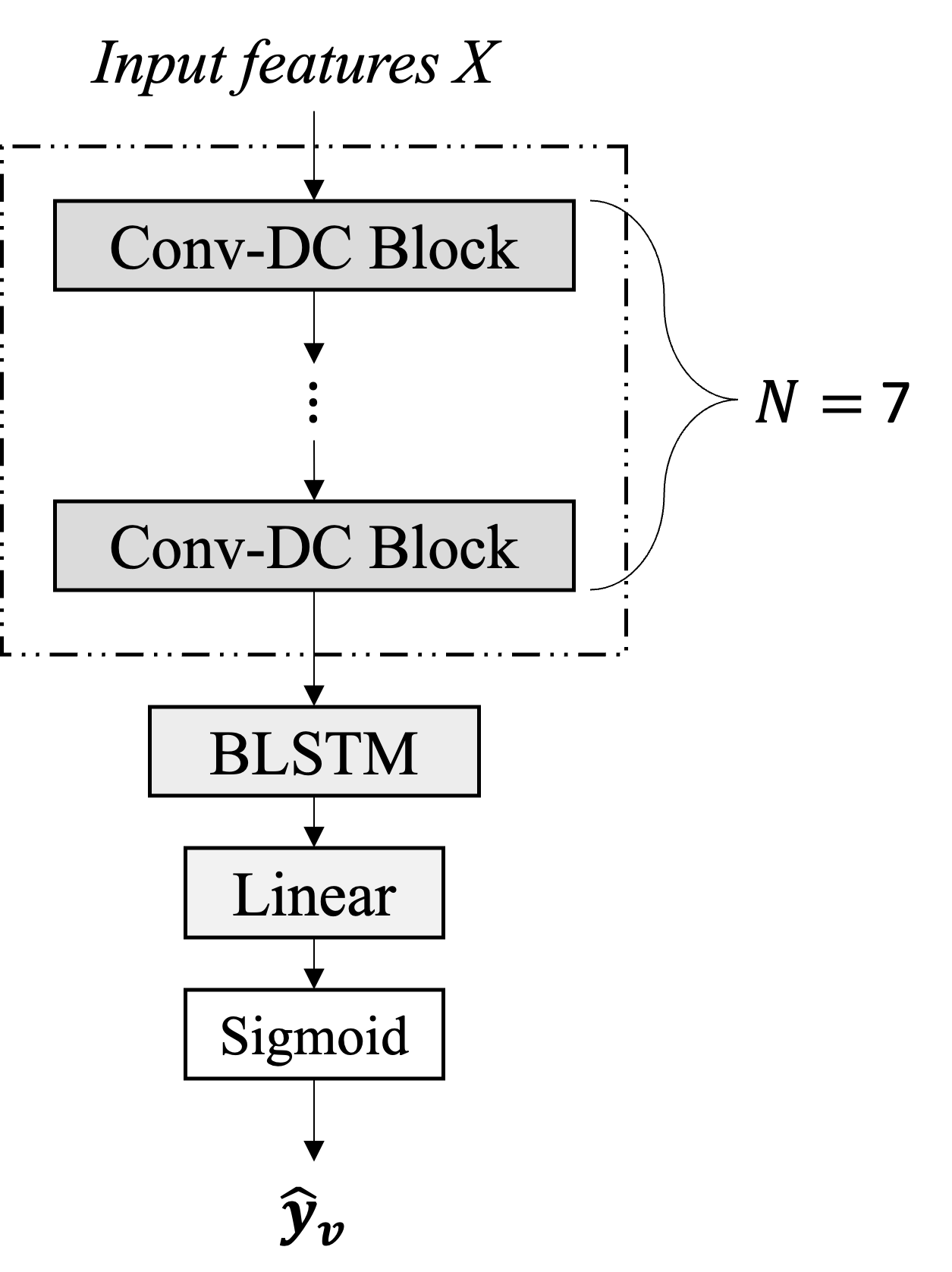}
  \caption{The proposed DC-CRN model for voicing detection, where $N$ is the number of Conv-DC blocks, and $\hat{y}_v$ represents the output for voicing detection and its value ranges from 0 to 1. }
  \label{fig:architecture}
\end{figure}

\begin{figure}[tp]
\centering
\begin{subfigure}[t]{0.8\linewidth}
    \centering
    \includegraphics[width=0.8 \linewidth]{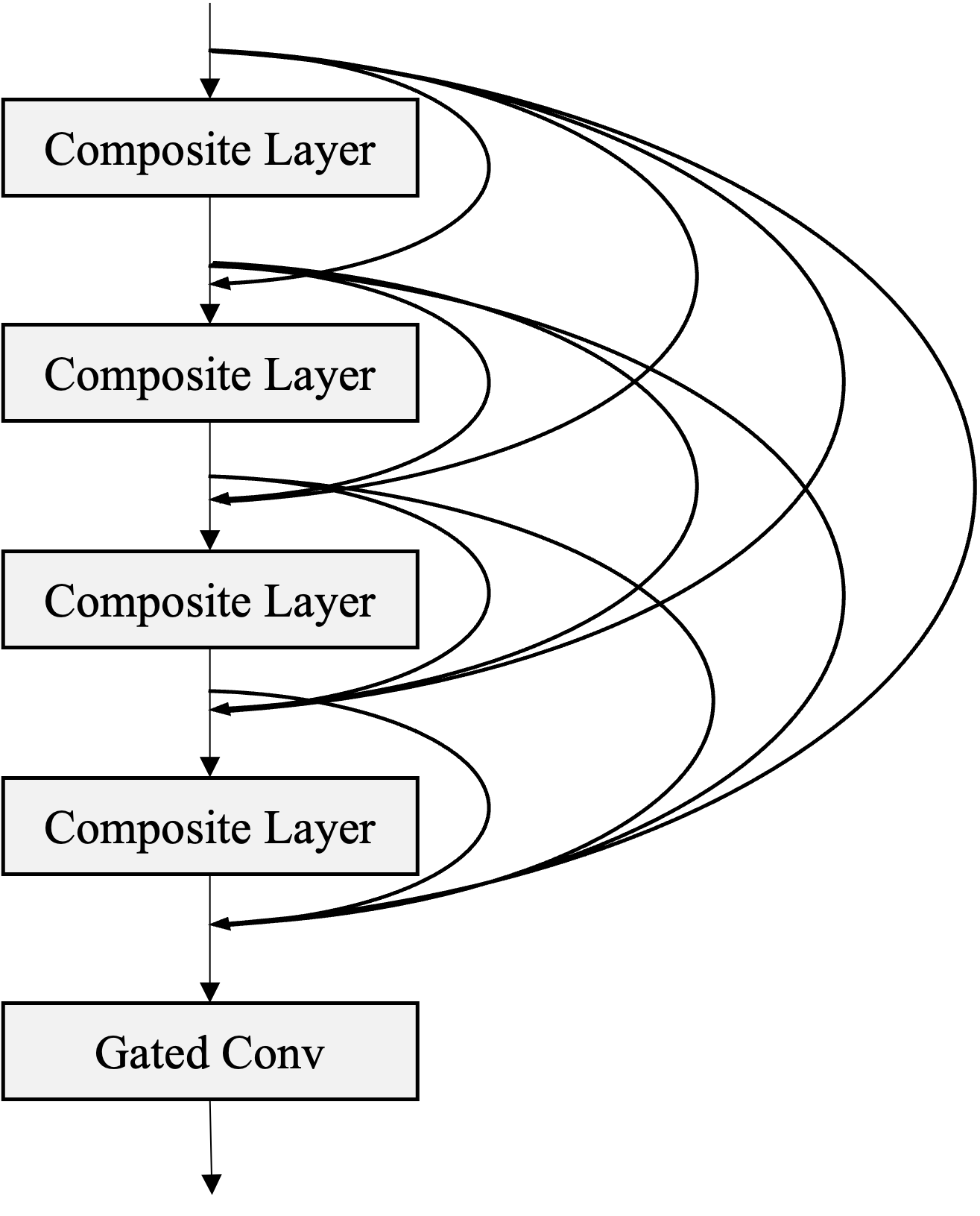}
    \caption[(a)]{Conv-DC Block}
\end{subfigure}\vfil
\vspace{3.99mm}
\begin{subfigure}[t]{0.9\linewidth}
    \centering
    \includegraphics[width=0.4\linewidth]{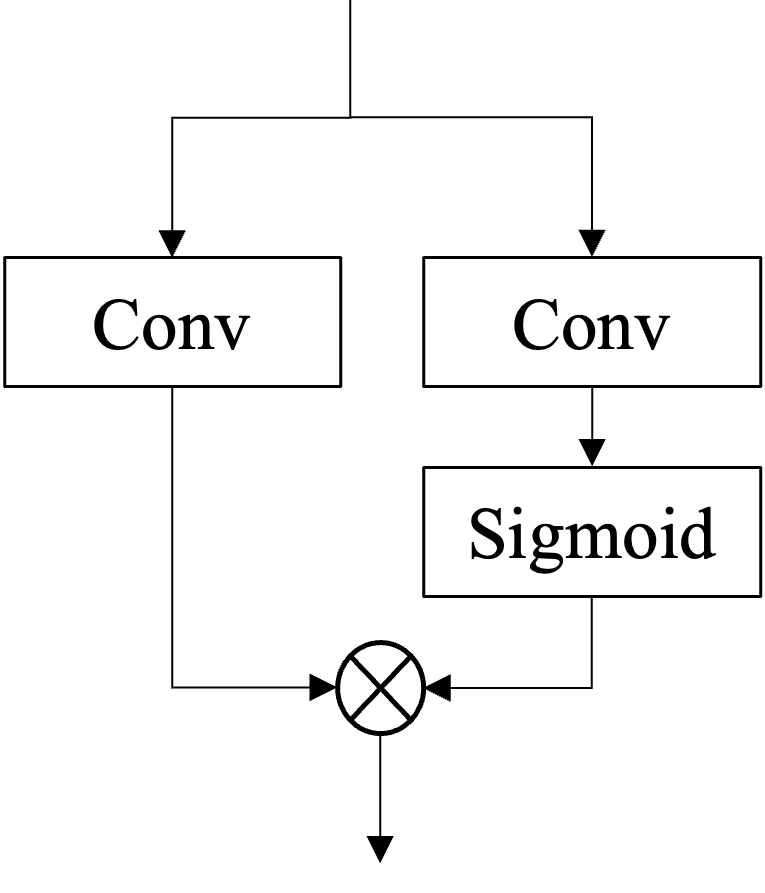}
    \caption[b]{Gated Convolution}
 \end{subfigure}
 \caption{Diagrams of (a) a DC-CRN block and (b) gated convolution. In (a), each composite layer contains a convolutional layer followed by batch normalization and ELU activation function. In (b), $\bigotimes$ denotes element-wise multiplication.}\label{fig:subblock}
\end{figure}

\begin{figure}[t]
  \centering
  \includegraphics[width=0.8\linewidth]{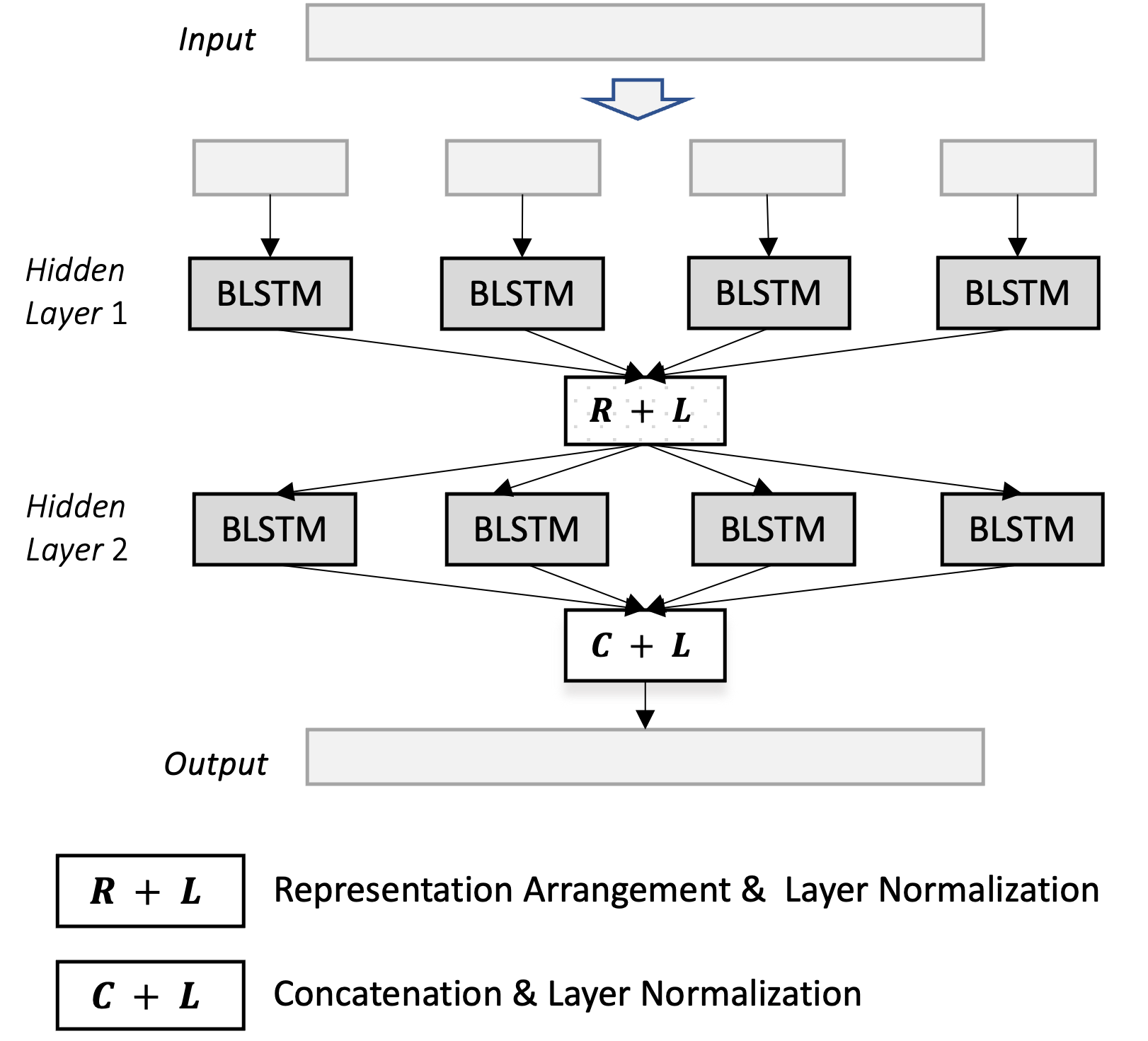}
  \caption{Grouping strategy for two-layer BLSTM, where the group number is set to 4.}
  \label{fig:group}
\end{figure}

In a previous study \cite{zhang2022F0}, inspired by the original DC-CRN model proposed for speech enhancement \cite{tan2021deep}, we developed a DC-CRN model for both fundamental frequency ($F0$) estimation and voicing detection in noisy speech. In this study, we adopt the same network architecture to tackle voicing detection in clean speech. 

The network structure is illustrated in Figure \ref{fig:architecture}. Following \cite{zhang2022F0} where better $F0$ estimation and voicing detection performance is obtained in the complex domain rather than in the magnitude domain, we choose a complex-domain input feature which is a concatenation of the real and imaginary parts of the complex short-time Fourier transform (STFT) of a speech signal,  
\begin{equation}
     X_{t,f} = [\Re(S_{t,f}), \Im(S_{t,f}])] \label{eq:input}
\end{equation}
where \( S_{t,f} \) represents the STFT of the speech signal at time \( t\) and frequency \( f\), $\Re$ and $\Im$ denote the real and imaginary parts, respectively. The advantages of complex-domain features over magnitude ones have been consistently shown for many speech processing tasks such as speech enhancement \cite{williamson2015complex} and speaker separation \cite{liu2019divide}. As shown in Sec. \ref{sec:eval}, we also find that complex-domain features lead to better voicing detection. Therefore, we adopt the complex-domain input feature in this study.

As shown in Figure \ref{fig:architecture}, the DC-CRN architecture consists of 7 convolutional densely-connected (Conv-DC) blocks, a two-layer bidirectional long short-term memory (BLSTM) block, and one linear layer followed by sigmoidal activation to produce the probabilistic output $\hat{y}_v$. A Conv-DC block in the DC-CRN network is shown in Figure \ref{fig:subblock}. A Conv-DC block comprises four composite layers followed by a gated convolutional layer. In each Conv-DC block, the outputs from all the preceding composite layers are concatenated to form the input to each composite layer or gated convolutional layer (see Figure \ref{fig:subblock}(a)). Specifically,
\begin{equation}
    c_l = H_l([c_{l-1},c_{l-2},...,c_0]), l=1,...,5,
\end{equation}
where $c_l = H_l(\cdot)$ is the output of the $l$th layer, $c_0$ is the input to the Conv-DC block, $H_l$ represents the mapping function performed by the $l$th composite layer, and $[...]$ denotes the concatenation operation. The dense connectivity allows a layer to reuse the features computed in all the preceding layers and results in better information flow through the layers. Each composite layer contains a 2D convolutional layer, followed by batch normalization and activation through the exponential linear unit (ELU). Figure \ref{fig:subblock}(b) depicts the final layer of the block, which is a gated convolutional layer that incorporates gated linear units introduced in \cite{dauphin2017language} and results in a masked convolution,
\begin{equation}
    \mathbf{v} = (\mathbf{u} * \mathbf{W}_1 + \mathbf{b}_1) \odot \sigma(\mathbf{u} * \mathbf{W}_2 + \mathbf{b}_2) = \mathbf{m}_1  \odot \sigma(\mathbf{m}_2)
\end{equation}
where $\mathbf{W_*}$, $\mathbf{b_*}$, and $\sigma(\cdot)$ represent kernels, biases, and the sigmoidal function, respectively. $\mathbf{u}$ and $\mathbf{v}$ represent the input and output of the gated convolutional block. $*$ and $\odot$ denote convolution and element-wise multiplication. The corresponding gradient is calculated as, 
\begin{equation}
    \nabla [\mathbf{m}_1 \odot \sigma(\mathbf{m}_2)] = \nabla \mathbf{m}_1 \odot \sigma(\mathbf{m}_2) + \sigma'(\mathbf{m}_2) \nabla \mathbf{m}_2 \odot \mathbf{m}_1
\end{equation}

To minimize the number of trainable parameters and improve computational efficiency, we follow \cite{tan2021deep} by employing a grouping technique originally proposed by Gao et al. \cite{gao2018efficient}. Figure \ref{fig:group} illustrates the grouping strategy for a two-layer BLSTM network with a group number of 4. We first split the input features and hidden states of the first two recurrent layers into 4 non-overlapping groups.
Intra-group features are learned within each group in the first recurrent layer. The forward and backward hidden states for the $j$th group, denoted as $\overrightarrow{\mathbf{h}}_{1,t}^j$ and $\overleftarrow{\mathbf{h}}_{1,t}^j$, are computed based on the input of the $j$th group $\mathbf{x}_t^j$ and the previous or subsequent hidden states $\overrightarrow{\mathbf{h}}_{1,t-1}^j$ and $\overleftarrow{\mathbf{h}}_{1,t+1}^j$, respectively,
\begin{equation}
\begin{aligned}
\overrightarrow{\mathbf{h}}_{1,t}^j &= f_{\text{forward}}^j(\mathbf{x}_t^j, \overrightarrow{\mathbf{h}}_{1,t-1}^j), \\
\overleftarrow{\mathbf{h}}_{1,t}^j &= f_{\text{backward}}^j(\mathbf{x}_t^j, \overleftarrow{\mathbf{h}}_{1,t+1}^j), \quad j = 1, \ldots, 4.
\end{aligned}
\end{equation}


In the second recurrent layer, inter-group dependency is modeled from the rearranged outputs from the first recurrent layer. A concatenation of the outputs from the second recurrent layer forms the final output. 
Layer normalization is incorporated after each recurrent layer. It has been observed that this technique reduces computational complexity while maintaining satisfactory performance.  

In terms of network configuration details, the seven Conv-DC blocks of the proposed model have output channels of 4, 8, 16, 32, 64, 128, and 256, respectively. Each convolutional layer in a Conv-DC block employs a $1 \times 3$ (time $\times$ frequency) kernel size, and has eight output channels. We apply zero padding of size one to both sides of the frequency dimension. The gated convolutional layer uses a kernel size of $1 \times 4$, a stride of two, and zero-padding of one on both sides of the frequency dimension. The two-layer BLSTM is composed of 512 units in each direction for each layer.

To train the DC-CRN model and obtain the probabilistic output $\hat{y}_v$ for an estimated voicing decision, we minimize the binary cross-entropy loss $\mathcal{L}_v$ for voicing detection. The loss function is defined as follows,
\begin{equation}
\mathcal{L}_{v}(y_v, \hat{y}_v) = -y_v\log{\hat{y}_v} - (1-y_v)\log{(1-\hat{y}_v)},
\label{bce}
\end{equation}
where $y_v$ represents the binary ground-truth voicing label, with $y_v = 1$ denoting a voiced frame, and 0 indicating otherwise.

To get voicing decisions during inference, $\hat{y}_v$ is compared against a threshold of 0.5. The frame is decided as voiced if $\hat{y}_v$ is higher than 0.5, and unvoiced otherwise.

\section{Experimental Setup}
\label{sec:setup}

\subsection{Datasets}
We train our model on five laryngograph datasets as described in Sec. \ref{sec:datasets}: PTDB-TUG, Mocha-TIMIT, FDA, KEELE, and CMU Arctic. The PTDB-TUG and Mocha-TIMIT datasets were pre-processed using the method mentioned in \ref{ssec:preprocessing}. Additionally, we utilize a dataset consisting of 50k utterances from the train-clean-360 subset of LibriSpeech \cite{panayotov2015librispeech} for pretraining. The labels for these utterances are extracted using the RAPT algorithm. All audio files are downsampled to 8 kHz. For STFT computation, we use a Hamming window of 128 ms duration with a 10 ms frame shift.

\subsection{Training Methodology}

To evaluate the performance of our approach, we employ a leave-one-corpus-out technique. Specifically, we divide the data from four of the five datasets into a training set, which comprises 90\% of the data, and a validation set comprising the other 10\%. The remaining dataset is used for testing or evaluation, and we repeat this process four times. By using this technique, we obtain a comprehensive assessment of the effectiveness of our method across multiple datasets, while minimizing the potential for bias and overfitting.

To enhance the generalizability of our trained model across different speakers and datasets, we employ a pretraining strategy. Specifically, we start with the model that has been trained on 50,000 microphone recordings from the LibriSpeech dataset. We use RAPT to generate pseudo voicing labels for this pretraining. By incorporating this pretraining on LibriSpeech utterances that are more than four times those of the combined laryngograph datasets, we aim to improve the overall performance of the model on unseen data and speakers.

All models are trained with the Adam optimizer, with an initial learning rate of 0.0005 which is  reduced by 50\% if the loss is not decreased for more than 5 consecutive epochs. Gradient clipping is applied with a maximum value of 5 to avoid gradient explosion. We set the maximum training epoch number to 80 and all models converge within this limit.

\subsection{Evaluation Metrics}

We evaluate the performance of voicing detection using Voicing Decision Error (VDE) which indicates the percentage of frames that are wrongly classified in terms of voicing:
\begin{equation}
\label{vde}
VDE = \frac{N_{p\rightarrow n} + N_{n\rightarrow p}}{N}, 
\end{equation}
where $N$ represents the total number of frames, $N_{p\rightarrow n}$ is the number of the voiced frames that are misclassified as non-voiced and $N_{n\rightarrow p}$ the number of non-voiced frames that are misclassified as voiced.

\subsection{Baselines}

Our evaluation includes quantitative comparisons against several strong baselines, including both signal processing and deep learning methods. 
For signal processing methods, we choose RAPT \cite{talkin1995robust} and SRH-Variant \cite{wang2022robust} which is an improved version of SRH \cite{drugman2019joint}. It has been shown \cite{koutrouvelis2015fast} that RAPT performs very well for clean speech, while SRH shows strong voicing detection performance for noisy speech.

For a deep learning baseline, we select a recent DNN-based approach called PENN \cite{morrison2023cross}, which is extended from the DNN methods of CREPE \cite{kim2018crepe}, FCN \cite{ardaillon2019fully}, DeepF0 \cite{singh2021deepf0} and estimates the periodicity of each speech frame to classify it as voiced or unvoiced. Different from prior methods, PENN proposes a novel entropy-based method for extracting per-frame signal periodicity, which significantly enhances the classification accuracy of voiced and unvoiced speech frames.

For implementation of baselines, we use the code provided in the Speech Signal Processing Toolkit (SPTK) python package \cite{sptktool} for RAPT, and the original code provided in \cite{wang2022robust} for SRH-Variant. For PENN, we use the default pretrained model provided in \cite{morrison2023cross}, which corresponds to FCNF0++ pretrained on MDB-stem-synth and PTDB-TUG datasets, with a selected unvoiced threshold of 0.25; see Sec. VI in \cite{morrison2023cross}. To ensure fair comparisons, we re-align the results from each baseline method for the lowest VDE.

\section{Evaluations and Comparisons}
\label{sec:eval}
\begin{figure}[htp]
\centering
\includegraphics[width=0.5\textwidth]{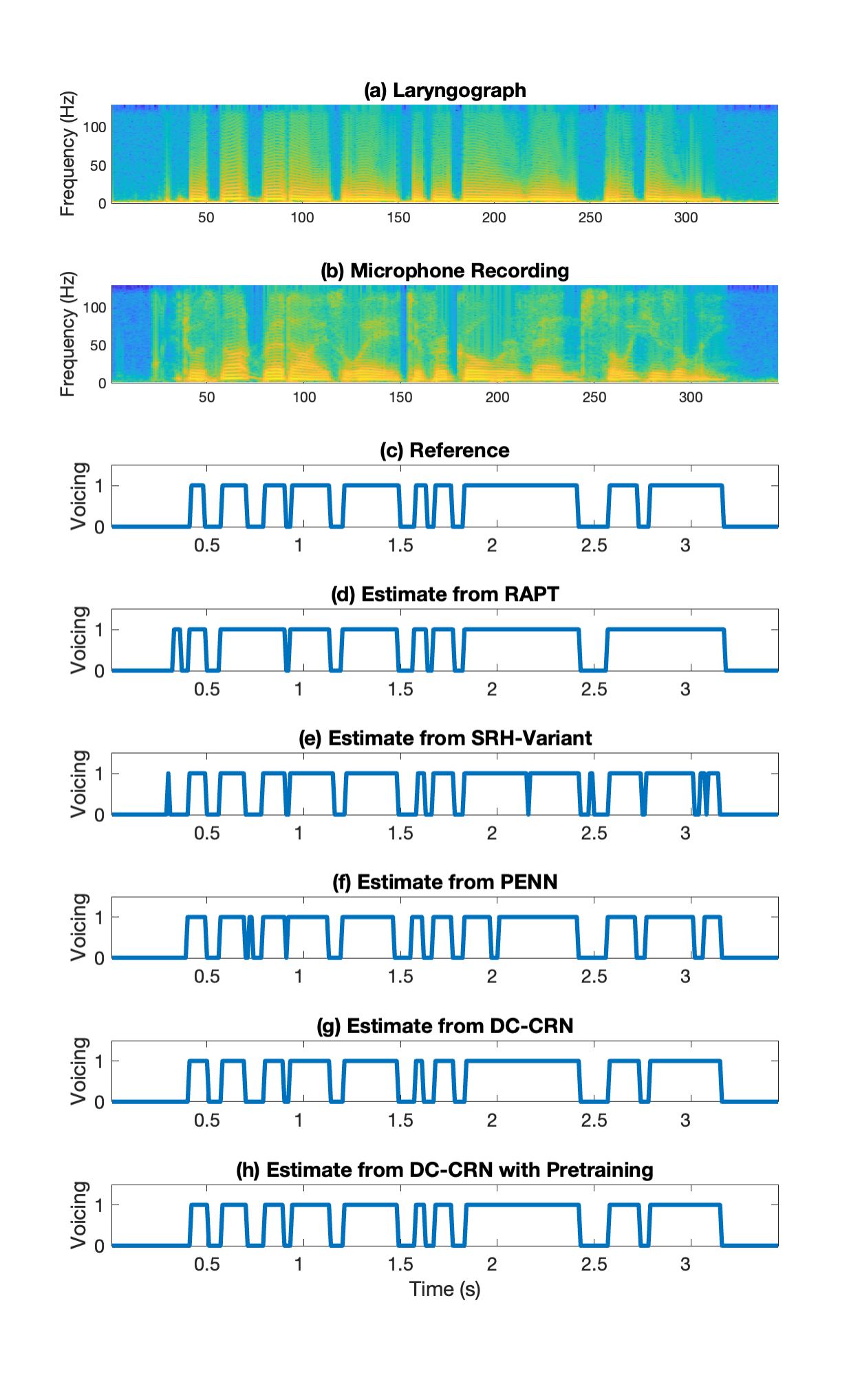}
\caption{An example of voicing detection in clean speech, which is a male utterance (``It was not Red-Eye's way to forego revenge so easily.") from the CMU Arctic corpus. (a) Spectrogram of laryngograph waveform, (b) Spectrogram of microphone recording, (c) Reference voicing decisions, (d) Estimated voicing decisions by RAPT, (e) Estimated voicing decisions by SRH-Variant, (f) Estimated voicing decisions by PENN, (g) Estimated voicing decisions by DC-CRN, (h) Estimated voicing decisions by DC-CRN with pretraining.}
\label{fig:example}
\end{figure}

\label{sec:eval}




\subsection{Cross Corpus Results}

\begin{table*}[]
\centering
\caption{Cross-corpus evaluation results in terms of voicing decision error (\%). The datasets used for training and evaluation are CMU Arctic (C), FDA (F), KEELE (K), Mocha-TIMIT (M), and PTDB-TUG (P).}
\label{tbl:crossvalidation}
\footnotesize
\begin{tabular}{l|c|cccccc}
\hline
\textbf{Training Set} & \textbf{Test Set}    & \textbf{RAPT}  & \textbf{SRH-Variant} & \textbf{PENN} & \textbf{DC-CRN (mag.)} & \textbf{DC-CRN} & \textbf{DC-CRN with pretraining} \\ \hline
M, K, F, C        & PTDB-TUG    & 3.47\% & 5.39\% & 2.37\% & 2.16\% & 2.03\% & \textbf{1.83\%}                   \\
P, K, F, C      & Mocha-TIMIT & 10.41\% & 13.53\% & 5.24\% & 3.96\% & 3.75\% & \textbf{3.43\%}                   \\
P, M, F, C         & KEELE       & 5.75\% & 9.37\% & 12.31\%& 5.49\% & 4.98\% & \textbf{4.06\%}                   \\
P, M, K, C       & FDA         & 4.61\% & 7.84\% & 10.05\% & 4.06\% & \textbf{3.77\%} & 4.32\%                   \\
P, M, F, K       & CMU Arctic  & 6.46\% & 8.79\% & 5.22\% & 5.04\% & 3.74\% & \textbf{3.47\%}                   \\ \hline
\end{tabular}
\end{table*}

To evaluate the proposed and baseline methods, we report a leave-one-corpus-out voicing detection results to assess cross-corpus generalization. The VDE results of the proposed methods and the baselines are given in Table \ref{tbl:crossvalidation}. It is worth noting that the PENN model is trained in part on PTDB-TUG, so some of the utterances in its PTDB-TUG evaluation are seen during training, resulting in potentially inflated results in this evaluation.

DC-CRN (mag.), DC-CRN,  and DC-CRN with pretraining, respectively, denote the versions of the proposed model with the magnitude STFT feature, complex STFT feature (i.e., Equation (\ref{eq:input})), and complex STFT feature with pretraining on LibriSpeech. Table \ref{tbl:crossvalidation} shows that Mocha-TIMIT is the most challenging dataset for the signal processing methods, with the VDE rates of 10.41\% for RAPT and 13.53\% for SRH-Variant. This may be attributed to the fact that the recorded speech signals in Mocha-TIMIT have significant device noise, making voicing detection more difficult. On the other hand, the deep learning methods show more tolerance to such noise. PENN cuts the VDE of RAPT by half, and DC-CRN with pretraining cuts the VDE by two thirds. It is worth noting that, as discussed in \cite{koutrouvelis2015fast}, RAPT has been shown to yield strong performance on clean speech, outperforming SRH-Variant on all datasets.

PENN, which was trained on a combination of PTDB and a synthetic dataset, outperforms the signal processing methods, except for the two small datasets of KEELE and FDA. In addition to the very low VDE on PTDB-TUG that is partly due to some common utterances in training and testing, on the Mocha-TIMIT dataset PENN achieves a VDE of 5.24\%, much better than 10.41\% by RAPT. On the CMU Arctic dataset, PENN achieves a VDE of 5.22\%, better than 6.46\% by RAPT. On the other hand, PENN performs poorly on the KEELE and FDA datasets, even worse than SRH-Variant, indicating a lack of generalization to these small datasets.

The proposed DC-CRN model produces the best results across all datasets. Better voicing detection performance is obtained in the complex domain than in the magnitude domain, consistent  with our recent observation in noisy speech \cite{zhang2022F0}. DC-CRN outperforms PENN on all datasets. On Mocha-TIMIT and CMU Arctic where PENN performs well, the DC-CRN model obtains the VDE of 3.75\% and 3.74\%, respectively, compared to PENN's 5.24\% and 5.22\%. On the small datasets where PENN does not perform well, the DC-CRN model achieves a VDE of 4.98\% for KEELE and 3.77\% for FDA, which are significantly lower than those from RAPT. These results suggest that the trained DC-CRN model have better generalization by leveraging multiple datasets. Furthermore, the DC-CRN model exhibits outstanding performance on the PTDB-TUG dataset, even surpassing that of the PENN model that is trained in part on this corpus. 

For the DC-CRN model pretrained on the LibriSpeech dataset, our evaluation results demonstrate consistent improvements across datasets with the exception of the small FDA corpus. For instance, pretraining improves the VDE on Mocha-TIMIT by 8.53\% and on Keele by 18.5\% relatively. These outcomes suggest that pretraining the DC-CRN model on LibriSpeech help to boost voicing detection performance across datasets.

Figure \ref{fig:example} illustrates voicing detection performed on a laryngograph recording, specifically a male utterance from the CMU Arctic corpus. Figure \ref{fig:example}(c) shows the reference voicing labels derived from Figure \ref{fig:example}(a) using the method described in Sec. \ref{ssec: label generation}. As shown in Figure \ref{fig:example}(d) and \ref{fig:example}(e), RAPT is prone to overestimating voiced regions, while SRH-Variant has both overestimation and underestimation errors. The DNN baseline, PENN, makes quality estimation but does not eliminate the underestimation problem. The proposed DC-CRN and DC-CRN with pretraining models show better voicing detection performance than PENN, yielding the most accurate estimate among all the methods.

\subsection{Cross Corpus Results on Provided Labels}

\begin{table*}[]
\centering
\caption{Cross-corpus evaluation results in terms of voicing decision error (\%) evaluated on datasets with provided labels. The datasets used for training and evaluation are CMU Arctic (C), FDA (F), KEELE (K), Mocha-TIMIT (M), and PTDB-TUG (P).}
\label{tbl:crossvalidationprovided}
\begin{tabular}{l|c|ccccc}
\hline
\textbf{Training Set}                             & \textbf{Test Set} & \textbf{RAPT}  & \textbf{SRH-Variant} & \textbf{PENN}   & \textbf{DC-CRN} & \textbf{DC-CRN with pretraining} \\ \hline
M, K, F, C      & PTDB-TUG & 4.29\% & 7.24\%            & 3.2\%   & 2.97\% &  \textbf{2.76\%}                        \\
P, M, F, C   & KEELE    & \textbf{4.52\%} & 7.84\%             & 13.53\% & 4.66\% &  4.55\%                        \\
P, M, K, C & FDA      & 4.85\% &  8.01\%           & 12.49\% & \textbf{4.73\%} &  5.34\%                        \\ \hline
\end{tabular}
\end{table*}

As explained before, three of the five laryngograph corpora provide voicing labels. We now evaluate the proposed and baseline methods using the provided labels, and the results are given in Table \ref{tbl:crossvalidationprovided}. It should be noted that, for the KEELE dataset, the provided labels include three classes: voiced, unvoiced, and uncertain. Our evaluation does not consider the uncertain frames for evaluation, which results in a lower VDE for KEELE compared to the results shown in Table \ref{tbl:crossvalidation}. It is observed that the two signal processing methods have comparable VDE rates across provided and generated labels, indicating the consistency and similarity of the labels generated using laryngograph data. 
In addition, the poor performance of PENN on the KEELE and FDA datasets confirms our earlier observation of its limited generalization ability. The VDE of the DC-CRN model is 1.32\% lower than that of RAPT on the PTDB-TUG dataset, and the results of DC-CRN and RAPT are close on the small KEELE and FDA datasets. These results demonstrate that the proposed DC-CRN model and
pretraining have strong generalizability across different label generation algorithms.

\section{Concluding Remarks}
\label{sec:conclusions}

This study introduces a robust DNN-based voicing detection model for clean speech by using laryngograph data for training. The model employs a DC-CRN architecture and incorporates a pretraining strategy on the LibriSpeech dataset. Our cross corpus evaluations demonstrate that the proposed model outperforms signal processing and deep learning baseline methods and shows strong generalization. By open sourcing the model and the data, we expect to accelerate the progress of voicing detection and related research such as pitch tracking in challenging environments.

Given the success of using synthesized speech data for training DNN models for pitch estimation (see, e.g., \cite{kim2018crepe}), should voicing detection utilize synthetic speech data?  
While synthetic data allows complete control of ground-truth voicing labels, our preliminary investigation suggests that using synthetic data is not effective for voicing detection. A possible reason is that a voicing detection model trained on synthetic data generated from microphone recordings relies heavily on the similarity of the synthetic training data and the voicing patterns of microphone recordings, which is difficult to maintain at the boundaries between voiced and unvoiced intervals. Also, voicing detection from the microphone recording of a speech utterance is prone to errors, as highlighted in this paper. On the other hand, laryngograph data represents the gold standard for $F0$ and voicing label generation. Prior studies on pitch estimation \cite{ardaillon2019fully, zhang2022densely} prefer synthetic data over laryngograph data partly because octave errors affect the accuracy of $F0$ labels derived from laryngograph recordings. This concern, however, does not extend to voicing labels.

In future work, we plan to apply the proposed voicing detection model to improve pitch tracking performance in adverse acoustic conditions, including background noise, room reverberation, and concurrent speakers.

\section*{Acknowledgments}
This work was supported in part by an NIDCD grant (R01DC012048) and the Ohio Supercomputer Center.

\bibliographystyle{IEEEbib}
\bibliography{strings,refs}

\begin{thebibliography}{10}

\bibitem{bai2021speaker}
Z. Bai and X.~L. Zhang,
\newblock ``Speaker recognition based on deep learning: An overview,''
\newblock {\em Neural Networks}, vol. 140, pp. 65--99, 2021.

\bibitem{wang2006computational}
D.~L. Wang and G.~J. Brown, Eds.,
\newblock {\em Computational auditory scene analysis: Principles, algorithms, and applications},
\newblock Hoboken, NJ: Wiley-IEEE Press, 2006.

\bibitem{zolnay2002robust}
A. Zolnay, R. Schl{\"u}ter, and H. Ney,
\newblock ``Robust speech recognition using a voiced-unvoiced feature,''
\newblock in {\em Proc. ICSLP}, 2002, pp. 1065--1068.

\bibitem{atal1976pattern}
B. Atal and L. Rabiner,
\newblock ``A pattern recognition approach to voiced-unvoiced-silence classification with applications to speech recognition,''
\newblock {\em IEEE Transactions on Acoustics, Speech, and Signal Processing}, vol. 24, pp. 201--212, 1976.

\bibitem{stevens1998acoustic}
K.~N. Stevens,
\newblock {\em Acoustic phonetics}, vol.~30,
\newblock Cambridge MA: MIT Press, 1998.

\bibitem{hu2008segregation}
G. Hu and D.~L. Wang,
\newblock ``Segregation of unvoiced speech from nonspeech interference,''
\newblock {\em Journal of the Acoustical Society of America}, vol. 124, pp. 1306--1319, 2008.

\bibitem{de2002yin}
A. De~Cheveign{\'e} and H. Kawahara,
\newblock ``Yin, a fundamental frequency estimator for speech and music,''
\newblock {\em Journal of the Acoustical Society of America}, vol. 111, pp. 1917--1930, 2002.

\bibitem{amado2008pitch}
R.~G. Amado and J. Vieira~Filho,
\newblock ``Pitch detection algorithms based on zero-cross rate and autocorrelation function for musical notes,''
\newblock in {\em Proc. ICALIP}, 2008, pp. 449--454.

\bibitem{bachu2010voiced}
R.~G. Bachu, S. Kopparthi, B. Adapa, and B.~D. Barkana,
\newblock ``Voiced/unvoiced decision for speech signals based on zero-crossing rate and energy,''
\newblock in {\em Advanced Techniques in Computing Sciences and Software Engineering}, 2010, pp. 279--282.

\bibitem{talkin1995robust}
D. Talkin and W.~B. Kleijn,
\newblock ``A robust algorithm for pitch tracking (\uppercase{RAPT}),''
\newblock {\em Speech Coding and Synthesis}, vol. 495, p. 518, 1995.

\bibitem{wang2022robust}
D. Wang, Y. Wei, Y. Wang, and J. Wang,
\newblock ``A robust and low computational cost pitch estimation method,''
\newblock {\em Sensors}, vol. 22, p. 6026, 2022.

\bibitem{han2014neural}
K. Han and D.~L. Wang,
\newblock ``Neural network based pitch tracking in very noisy speech,''
\newblock {\em IEEE/ACM Transactions on Audio, Speech, and Language Processing}, vol. 22, pp. 2158--2168, 2014.

\bibitem{tran2020robust}
D.~N. Tran, U. Batricevic, and K. Koishida,
\newblock ``Robust pitch regression with voiced/unvoiced classification in nonstationary noise environments.,''
\newblock in {\em Proc. Interspeech}, 2020, pp. 175--179.

\bibitem{zhang2022densely}
Y. Zhang, H. Wang, and D.~L. Wang,
\newblock ``Densely-connected convolutional recurrent network for fundamental frequency estimation in noisy speech,''
\newblock in {\em Proc. Interspeech}, 2022, pp. 401--405.

\bibitem{morrison2023cross}
M. Morrison, C. Hsieh, N. Pruyne, and B. Pardo,
\newblock ``Cross-domain neural pitch and periodicity estimation,''
\newblock {\em arXiv preprint arXiv:2301.12258}, 2023.

\bibitem{kim2018crepe}
J.~W. Kim, J. Salamon, P. Li, and J.~P. Bello,
\newblock ``Crepe: A convolutional representation for pitch estimation,''
\newblock in {\em Proc. ICASSP}, 2018, pp. 161--165.

\bibitem{panayotov2015librispeech}
V. Panayotov, G. Chen, D. Povey, and S. Khudanpur,
\newblock ``Librispeech: an \uppercase{asr} corpus based on public domain audio books,''
\newblock in {\em Proc. ICASSP}, 2015, pp. 5206--5210.

\bibitem{ladefoged2001vowels}
P. Ladefoged,
\newblock ``Vowels and consonants,''
\newblock {\em Oxford U.K.: Blackwell}, vol. 58, pp. 211--212, 2001.

\bibitem{bagshaw1993enhanced}
P.~C. Bagshaw, S.~M. Hiller, and M.~A. Jack,
\newblock ``Enhanced pitch tracking and the processing of f0 contours for computer aided intonation teaching,''
\newblock in {\em Proc. Eurospeech}, 1993, pp. 1003--1006.

\bibitem{drugman2019joint}
T. Drugman and A. Alwan,
\newblock ``Joint robust voicing detection and pitch estimation based on residual harmonics,''
\newblock in {\em Proc. Interspeech}, 2011.

\bibitem{narendra2015robust}
N. Narendra and K.~S. Rao,
\newblock ``Robust voicing detection and \uppercase{F}0 estimation for \uppercase{HMM}-based speech synthesis,''
\newblock {\em Circuits, Systems, and Signal Processing}, vol. 34, pp. 2597--2619, 2015.

\bibitem{koutrouvelis2015fast}
A.~I. Koutrouvelis, G.~P. Kafentzis, N.~D. Gaubitch, and R. Heusdens,
\newblock ``A fast method for high-resolution voiced/unvoiced detection and glottal closure/opening instant estimation of speech,''
\newblock {\em IEEE/ACM Transactions on Audio, Speech, and Language Processing}, vol. 24, pp. 316--328, 2015.

\bibitem{upadhyay2015instantaneous}
A. Upadhyay and R.~B. Pachori,
\newblock ``Instantaneous voiced/non-voiced detection in speech signals based on variational mode decomposition,''
\newblock {\em Journal of the Franklin Institute}, vol. 352, pp. 2679--2707, 2015.

\bibitem{kumar2016voice}
S.~S. Kumar and K.~S. Rao,
\newblock ``Voice/non-voice detection using phase of zero frequency filtered speech signal,''
\newblock {\em Speech Communication}, pp. 90--103, 2016.

\bibitem{mcaulay1990pitch}
R.~J. McAulay and T.~F. Quatieri,
\newblock ``Pitch estimation and voicing detection based on a sinusoidal speech model,''
\newblock in {\em Proc. ICASSP}, 1990, pp. 249--252.

\bibitem{atal1972automatic}
B.~S. Atal,
\newblock ``Automatic speaker recognition based on pitch contours,''
\newblock {\em Journal of the Acoustical Society of America}, vol. 52, pp. 1687--1697, 1972.

\bibitem{liu2018permutation}
Y. Liu and D.~L. Wang,
\newblock ``Permutation invariant training for speaker-independent multi-pitch tracking,''
\newblock in {\em Proc. ICASSP}, 2018, pp. 5594--5598.

\bibitem{plante1995pitch}
F. Plante, G. Meyer, and W. Ainsworth,
\newblock ``A pitch extraction reference database,''
\newblock in {\em Proc. Eurospeech}, 1995, pp. 837--840.

\bibitem{drugman2018traditional}
T. Drugman, G. Huybrechts, V. Klimkov, and A. Moinet,
\newblock ``Traditional machine learning for pitch detection,''
\newblock {\em IEEE Signal Processing Letters}, vol. 25, pp. 1745--1749, 2018.

\bibitem{kominek2004cmu}
J. Kominek and A.~W. Black,
\newblock ``The \uppercase{CMU} arctic speech databases,''
\newblock in {\em Fifth ISCA workshop on speech synthesis}, 2004, pp. 223--224.

\bibitem{pirker2011pitch}
G. Pirker, M. Wohlmayr, S. Petrik, and F. Pernkopf,
\newblock ``A pitch tracking corpus with evaluation on multipitch tracking scenario,''
\newblock in {\em Proc. Interspeech}, 2011, pp. 1509--1512.

\bibitem{zhang2022F0}
Y. Zhang, H. Wang, and D.~L. Wang,
\newblock ``\uppercase{$F0$} estimation and voicing detection with cascade architecture in noisy speech,''
\newblock {\em IEEE/ACM Transactions on Audio, Speech, and Language Processing}, vol. 31, pp. 3760--3770, 2023.

\bibitem{tan2021deep}
K. Tan, X. Zhang, and D.~L. Wang,
\newblock ``Deep learning based real-time speech enhancement for dual-microphone mobile phones,''
\newblock {\em IEEE/ACM Transactions on Audio, Speech, and Language Processing}, vol. 29, pp. 1853--1863, 2021.

\bibitem{williamson2015complex}
D.~S. Williamson, Y. Wang, and D.~L. Wang,
\newblock ``Complex ratio masking for monaural speech separation,''
\newblock {\em IEEE/ACM Transactions on Audio, Speech, and Language processing}, vol. 24, pp. 483--492, 2016.

\bibitem{liu2019divide}
Y. Liu and D.~L. Wang,
\newblock ``Divide and conquer: A deep \uppercase{CASA} approach to talker-independent monaural speaker separation,''
\newblock {\em IEEE Transactions on Audio, Speech, and Language Processing}, vol. 27, pp. 2092--2102, 2019.

\bibitem{dauphin2017language}
Y.~N. Dauphin, A. Fan, M. Auli, and D. Grangier,
\newblock ``Language modeling with gated convolutional networks,''
\newblock in {\em Proc. 34th Int. Conf. Mach. Learn.}, 2017, pp. 933--941.

\bibitem{gao2018efficient}
F. Gao, L. Wu, L. Zhao, T. Qin, X. Cheng, and T.-Y. Liu,
\newblock ``Efficient sequence learning with group recurrent networks,''
\newblock in {\em Proc. Conference of the North American Chapter of the Association for Computational Linguistics: Human Language Technologies}, 2018, pp. 799--808.

\bibitem{ardaillon2019fully}
L. Ardaillon and A. Roebel,
\newblock ``Fully-convolutional network for pitch estimation of speech signals,''
\newblock in {\em Proc. Interspeech}, 2019, pp. 2005--2009.

\bibitem{singh2021deepf0}
S. Singh, R. Wang, and Y. Qiu,
\newblock ``Deep\uppercase{F}0: End-to-end fundamental frequency estimation for music and speech signals,''
\newblock in {\em Proc. ICASSP}, 2021, pp. 61--65.

\bibitem{sptktool}
\uppercase{SPTK}~working group,
\newblock ``Speech signal processing toolkit (\uppercase{SPTK}),'' \url{https://sp-tk.sourceforge.net/}.

\end{thebibliography}

\end{document}